\documentclass[%
 reprint,
longbibliography,
 amsmath,amssymb,
 aps,
pra
]{revtex4-2}

\usepackage{nameref}
\usepackage{units}
\usepackage{mathtools}
\usepackage{graphicx}
\usepackage{dcolumn}
\usepackage{bm}
\usepackage{hyperref}
\usepackage[dvipsnames]{xcolor}
\usepackage{soul}
\usepackage{lineno}

\usepackage{braket}
\usepackage{physics}
\usepackage{appendix}
\usepackage{float}
\usepackage{wasysym}
\usepackage{subeqnarray}
\usepackage{verbatim}
\usepackage{marginnote}
\usepackage{xfrac}
\usepackage{makecell}
\usepackage{placeins}

\newcommand{\Er}{\, E_{\rm r}}

\begin{document}
\title{Probing the Rung‑Mott Insulator in Bosonic Two‑Leg Ladders}

\author{Christopher Parsonage$^\ast$, Lennart Koehn$^\ast$, Lorenzo Carfora, Ben Watson, Timon Hilker, Elmar Haller, Peter Kirton, Arthur La Rooij, Stefan Kuhr}

\address{University of Strathclyde, Department of Physics, Glasgow G4 0NG, United Kingdom\\
\smallskip
\small$^\ast$\rm{These authors contributed equally to this work.}
}

\date{\today}


\date{\today}

\begin{abstract}

Mott insulators typically appear at integer densities, but certain lattice geometries or non-local interactions can stabilise Mott insulators at fractional densities. In this work, we study strongly interacting two-leg ladders using ultracold bosonic atoms around half-filling, and identify the rung-Mott insulating (RMI) phase through its characteristic rung occupation in site-resolved measurements. We probe its excitation spectrum via lattice modulation spectroscopy, and interpret the observed spectra using  numerical simulations that include the driven dynamics, coupling to multiple excited states, and energy offsets. As with other Mott insulators, the RMI phase is characterized through its incompressibility, which we probe via a site-resolved analysis of densities and chemical potentials. Our results provide direct insight into the excitation dynamics of a strongly correlated ladder system, and pave the way for studying transport and further dynamical properties in low-dimensional quantum systems with programmable lattice geometries.
\end{abstract}

\maketitle

In quantum many-body systems realized in optical lattices, the geometry and dimensionality  strongly influence which  phases and dynamics occur~\cite{Lewenstein2012,Kimura1997,Hirthe2023}. In one dimension, quantum fluctuations prevent true long-range order in the weakly interacting regime~\cite{Cazalilla2011}. Instead, bosons form a quasi-condensate with algebraically decaying correlations, describing a Luttinger liquid. For strong repulsion, the bosons fermionize and and approach the strongly correlated Tonks-Girardeau regime~\cite{Paredes2004,Kinoshita2004}. In a lattice, the hard-core boson limit remains gapless at incommensurate filling, whereas at commensurate filling a Mott-insulating phase can emerge. 
In a lattice, this corresponds to hard-core bosons at incommensurate filling, while at commensurate filling the system becomes a Mott insulator.
Coupling two such chains  to form a ladder geometry supports phases and transport phenomena absent in a single chain~\cite{dagotto1992superconductivity,dagotto1996surprises,deshpande2010electron}. Realized in a variety of solid-state and molecular materials, ladder geometries exhibit transport and magnetic properties that depend strongly on the competition between leg and rung couplings. This interplay gives rise to effects including spin-gapped phases, correlated insulating behavior, and unconventional superconductivity ~\cite{abbamonte2004crystallization,ogura2017,fradkin2015colloquium,silva2021spin}.

A two-leg ladder consists of two one-dimensional chains (the legs) coupled by transverse links (the rungs), allowing particles to tunnel both along each leg and between the legs. For bosons on such a ladder, the interplay between on-site interactions, leg and rung tunnelling gives rise to a rich phase diagram. At commensurate filling, increasing the interactions drives the superfluid-to-Mott-insulator transition~\cite{Donohue2001,LuthraMishraPaiDas2008} that now depends on the two separate tunnelling terms. For incommensurate filling, such ladder systems support an additional insulating phase, called the rung-Mott insulator (RMI)~\cite{Crepin2011,Carrasquilla2011,Padhan2023,Padhan2024_SSHLadder,Carfora2026mott}. In this phase, particles delocalize across individual rungs while remaining localized along the leg direction of the ladder, resulting in an incompressible insulating state at fractional filling. In the strongly interacting limit, bosonic ladders can be mapped onto effective spin-ladder Hamiltonians~\cite{Crepin2011}, where the sites occupied by an atom act as one spin, and empty sites (holes) as the other spin. The RMI phase therefore provides a bosonic analogue of low-dimensional quantum magnetic systems and for studying strongly correlated transport~\cite{nichols2019spin}. Such analogies are particularly appealing in ultracold-atom platforms, where both local occupations and dynamical processes can be measured directly. 

\begin{figure}[b]
    \centering
    \includegraphics[width=0.95\linewidth]{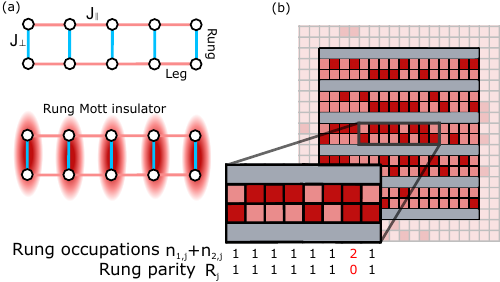}
    \caption{\textbf{Two-leg ladder systems.} (a) Two-leg ladder geometry, with tunneling amplitudes  along each leg $J_{\parallel}$ (red links) and along the rungs $J_\bot$ (blue links). In the rung-Mott insulating   phase, atoms are delocalised across the rungs, while tunneling along the legs is suppressed by interactions.
    (b) Reconstructed atom occupations ($V_\bot=5\Er$, $V_\parallel=20\Er$, $J_{\bot}/J_{\parallel}=27$, $U/J_\parallel=109$). Red and pink squares represent parity-projected lattice site occupations, with one and zero atoms, respectively.  Repulsive light potentials (grey shaded regions)  separate the system into five independent two-leg ladders. The magnified view  shows the RMI phase emerging at the centre, with rung occupations and rung parity for each rung.
    }
    \label{fig1}
\end{figure}

Earlier theoretical work explored the emergence of the RMI phase and the regimes in which bosonic and fermionic statistics differ, particularly in the presence of disorder~\cite{Crepin2011,Carrasquilla2011,block2011exotic}. More recently, finite-interaction studies have explored how the extent of the RMI phase depends on both tunnelling amplitudes and the interaction strength~\cite{Carfora2026mott}. Experimentally, ultracold atoms in optical lattices provide a highly controllable platform for realizing such ladder Hamiltonians~\cite{Aidelsburger2011,Tai2017,sompet2022realizing}. Previous experiments have observed chiral currents and Meissner physics in bosonic ladders~\cite{Atala2014},
and more recently reported the role of strong interactions~\cite{Impertro2025}, for which  enhanced density anti-correlations between the rung sites have been observed. Recent experimental work on topological criticality also showed the opening of an energy gap in related systems with two coupled chains \cite{Su2026}.

In this work, we realise coupled two-leg bosonic ladders in an optical lattice using programmable light potentials. We identify the RMI phase at  finite interaction strength via measurements of the rung parity as a function of filling and tunnelling strength, and further characterize the system  through probing of its excitation gap using amplitude-modulation spectroscopy. We also  determine the compressibility from on-site occupations as a function of local chemical potential.

Our experiments employ a quantum-gas microscope~\cite{Bakr2009,Haller2015,Cheuk2015,Buob2024,Kuhr2016}, allowing for single-site-resolved atom imaging. In each  realisation of the experiment, we prepare $^{87}$Rb atoms in five independent two-leg ladder systems, each comprising $2\times 16$ sites, between six parallel repulsive potential barriers [Fig.~\ref{fig1}(b)]. The  system is described by the Bose-Hubbard model 
\begin{eqnarray}
    \hat{H} &=& -J_{\parallel} \sum_{i,j} (\hat{a}_{i,j}^\dagger \hat{a}_{i,j+1} + {\rm h.c.})-J_{\bot}  \sum_{j} (\hat{a}_{1,j}^\dagger \hat{a}_{2,j} +{\rm h.c.}) \nonumber\\
     &&+ \sum_{i,j} \frac{U}{2} \hat{n}_{i,j}(\hat{n}_{i,j}-1) 
     + \sum_{i,j}(\epsilon_{i,j}  - \mu)\hat{n}_{i,j} \,,
\label{eq:BHM}
 \end{eqnarray}
where $J_{\parallel}$ and $J_{\bot}$ denote the tunnelling amplitudes along the legs and rungs, respectively [Fig.~\ref{fig1}(a)]. Here, $\hat{a}_{i,j}^\dagger$ ($\hat{a}_{i,j}$) is the bosonic creation (annihilation) operator with rung and leg indices $j=1,...,N$, and $i=1,2$ respectively, and  $\hat{n}_{i,j}=\hat{a}_{i,j}^\dagger\hat{a}_{i,j}$ is the corresponding number operator. $U$ is the on-site interaction energy, $\epsilon_{i,j}$ is a local potential offset, and $\mu$ is the chemical potential. The competition between tunnelling and interactions in Eq.~\ref{eq:BHM} gives rise to superfluid as well as insulating phases. In the weakly interacting regime, when $J_{\bot},J_{\parallel} > U$, atoms are delocalized across the ladder and form a superfluid state. At interaction energies $U \gg J_{\bot},J_{\parallel}$, a Mott-insulating phase with energy gap $\Delta E=U$ can form at unit filling. For weak coupling between the legs, $J_\bot \ll J_\parallel$, the system separates into nearly isolated 1D systems. By contrast, when the rung coupling becomes larger than the  coupling within the legs, $J_\bot > J_\parallel$, the system can enter the RMI phase at half filling. A single atom is delocalized across each rung, while transport along leg direction is suppressed. This state is insulating and exhibits a finite energy gap $\Delta E < U$. To identify the RMI phase, we determine the rung occupations and define the rung parity as $R_j = {\rm mod}_2(n_{1,j}+n_{2,j})$, where $n_{i,j}$ denotes the occupation of rung $j$ on leg $i$ [Fig.~\ref{fig1}]. In the RMI phase, each rung is occupied by a single atom, yielding $R_j = 1$, whereas empty rungs or rungs containing two atoms give $R_j=0$. The average rung parity for a ladder with $N$ rungs is defined as $R=\sum_{j=1}^N R_j/N$. For a superfluid above half-filling, the parity-projected site occupations are $\sim0.5$ on average, resulting in $R\simeq 0.5$.

\begin{figure}[!b]
    \centering
    \includegraphics[]{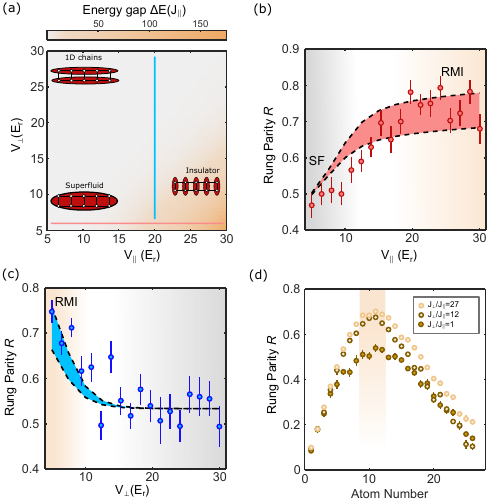}
    \caption{\textbf{Observation of the RMI phase.} (a)  Numerical calculation of the energy gap, $\Delta E$, of the half-filled ladder as a function of lattice depths $V_{\bot}$, $V_{\parallel}$. For small $V_{\parallel}$,  coupled or decoupled 1D superfluid states are present, whereas for $V_{\bot} \ll  V_{\parallel}$ the  RMI phase emerges (brown shaded region). The red and blue lines show the parameter scans used for the datasets in (b) and (c).  (b)  Rung parity $R$ for increasing $V_\parallel$ ($V_\bot=5\Er$). The shaded region shows the result from DMRG calculations for a range of temperatures, 
       $10.1 J_\parallel < k_BT < 13.5 J_{\parallel}$ for $V_\parallel=20\Er$.
       (c) $R$ for increasing $V_\bot$ ($V_\parallel=20\Er$) together with numerical  calculations (blue shaded region) for the same temperatures as in (b). (d) 
    Rung parity vs total atom number in the $2 \times 16$ site ladder. The brown-shaded region shows the range of detected atom numbers (9-12 atoms) we use to post-select our datasets for the results shown in (b) and (c). The lattice depths for the three datasets are $V_\bot=5\Er,V_\parallel=20\Er$ for $J_\perp/J_\parallel=27$, $V_\bot=8\Er,V_\parallel=20\Er$ for $J_\perp/J_\parallel=12$, and $V_{\bot,\parallel}=20\Er$ for $J_\perp/J_\parallel=1$. Error bars shown in b, c and d are statistical. Some error bars are smaller than the marker size.}
    \label{fig:2}
\end{figure}

In a first set of experiments, we investigate  the emergence of the RMI phase by measuring the rung parity for different regimes of $J_{\bot}/J_{\parallel}$. To identify the parameter region where  the RMI is expected  to occur  at half-filling, we use density-matrix renormalisation group (DMRG) calculations~\cite{Carfora2026mott} (Supplemental Material) to obtain the energy gap over a range of lattice depths $V_\bot$ and $V_\parallel$ [Fig.~\ref{fig:2}(a)]. Guided by these  predictions, we vary $V_\parallel$ from $5\Er$ to $30\Er$, while keeping  $V_\bot=5\Er$ fixed. For this parameter range, we observe an increase of the rung parity from $R\simeq0.5$ to $R\simeq0.75$, indicating a transition from a superfluid regime towards the RMI phase [Fig.~\ref{fig:2}(b)]. Due to the finite system size and temperature, we identify this as a crossover rather than a sharp transition \cite{Carfora2026mott}. The maximum observed rung parity is limited by finite temperature effects, consistent with our numerical simulations. By comparing  the maximum observed  $R_p$ in the RMI regime (V$_{\parallel}=20 \Er$, $V_{\perp}=5 \Er$)  we infer a temperature  $k_BT\approx12(2)J_{\parallel}$, corresponding to the centre of the uncertainty band in Fig.~\ref{fig:2}(b). This temperature is smaller than the gap energy, $\Delta E=27  J_{\parallel}$.  In a second experiment, we begin in  the RMI regime  ($V_\bot=5\Er, V_\parallel=20\Er$) and increase $V_\bot$. We observe a reduction of the rung parity from $R\simeq0.75$ to $R\simeq0.5$ [Fig.~\ref{fig:2}(c)], again in agreement with the finite temperature numerical calculations. Here, the interaction energy becomes  too large  for the system to enter a  superfluid phase state at our finite temperatures, while the density remains too low for the formation of a conventional  Mott insulator.

The harmonic confinement of the trap results in  a spatially varying density across the ladder. A higher atom number reduces the measured rung parity as the energy gap of the RMI is less than the energy offset from the harmonic confinement away from the trap centre. To characterise this effect, we calculate  the average rung parity over the central eight rungs as a function of  atom number for three different lattice depths, allowing us  to identify the parameter range  where the RMI forms [Fig.~\ref{fig:2}(d)]. We find  the highest rung parity for detected atom numbers between 9 to 12, and had used this criterion to post-select the datasets presented in Figs.~\ref{fig:2}(b) and~\ref{fig:2}(c). 
For higher atom numbers, we observe that the rung parity decreases towards  zero, consistent with the formation of the Mott insulator. 

\begin{figure}
    \centering
    \includegraphics[width=1\linewidth]{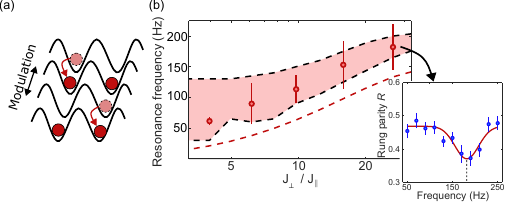}
    \caption{\textbf{Modulation resonance.} (a) 
    Illustration of induced tunnelling in a modulated lattice,  where an atom tunnels to a neighboring rung already occupied by one atom. (b) Measured  resonance frequencies for $J_{\perp} / J_{\parallel} = 4,6,10,16,27$. The datasets were post-selected to include a range of $6-16$ atoms on the 16 rungs, because the modulation causes a larger variation of atom numbers. Error bars are the width of the resonance. 
    The red shaded region shows  the numerically calculated resonance frequencies, for a potential bias $\epsilon_B$ between the two legs ($0 < \epsilon_B/J_\perp < 1.1$) (Supplemental Materials). The red dashed line shows the energy gap calculated with the DMRG method, at $\epsilon_B=0$. Inset: Rung parity $R$ as function of modulation frequency, with statistical error bars. The red line is a Gaussian fit to determine the  resonance centre.
    }
        \label{fig:3}
\end{figure}

To probe the excitation spectrum of the RMI, we  modulate the lattice depth $V_\parallel$ \cite{ma2011photon} to induce tunnelling of an  atom across a rung to a neighboring rung that is already occupied by one atom [Fig.~\ref{fig:3}(a)]. This process creates additional  empty and doubly occupied rungs, resulting in a reduction of  the measured rung parity $R$. We modulate the lattice sinusoidally with a modulation amplitude of 25\% for durations between $30$ and $100$\,ms, across a range of values $J_\bot/J_\parallel$ [Fig.~\ref{fig:3}(b)]. For each parameter set, we measure $R$ as a function of the modulation frequency, and identify the  resonance frequency from the position of the minimum in the rung parity [inset in Fig.~\ref{fig:3}(b)].  We find that the measured resonance frequency increases with $J_\bot/J_\parallel$, but lies systematically above the excitation  gap predicted by the DMRG calculations [red line in Fig.~\ref{fig:3}(b)]. To understand this discrepancy, 
we use a time-dependent $V_{\parallel}(t)$ to capture the effect of the modulation used in the experiment. We further include the local energy offsets $\epsilon_{i,j}=\epsilon'_{i,j}+\epsilon_B$, where $\epsilon'_{i,j}$ originates form the harmonic confinement and $\epsilon_B$ describes a relative energy offset between the two legs due to  experimental imperfections (Supplemental Material).

When we simulate only the modulation of $V_{\parallel}(t)$ and neglect the local energy offsets ($\epsilon_{i,j}=0$), we already observe a shift of the resonance to higher frequencies, as indicated by the lower-bound black line in Fig.~\ref{fig:3}(b). This difference is primarily caused by power broadening due to the finite  modulation amplitude. The lattice modulation couples the RMI ground state  to multiple  excited states with different transition matrix elements (Supplemental Material~\ref{Suppl:modulation}). For a finite ladder, several of these transitions are accessible, and  the observed spectral response is determined by both the energy separation of the excited states and their coupling strength to the drive. At weak modulation amplitudes, these individual resonances could in principle be resolved, whereas for the stronger modulation used experimentally they overlap and merge into a single broadened feature. As a result, the resonance is shifted towards the transitions with the largest coupling strengths, rather than occurring exactly at the  many-body excitation gap obtained without including the modulation. 

Including the leg-dependent potential bias $\epsilon_B$ leads to a further shift of the  resonance, bringing the simulations in good agreement with the experimental data for $0 < \epsilon_B/J_{\perp} < 1.1$. The upper value of $\epsilon_B$  corresponds to an energy offset equal to $\sim$~5\% of the repulsive light potential on the neighboring site ($\sim 3\,U$ in the RMI regime). Such a bias can  arise from  fluctuations in the position of the potential barriers and from spatial  inhomogeneities in the intensity of the light potential (Supplemental Material).
\begin{figure}
\includegraphics[width=1\linewidth]{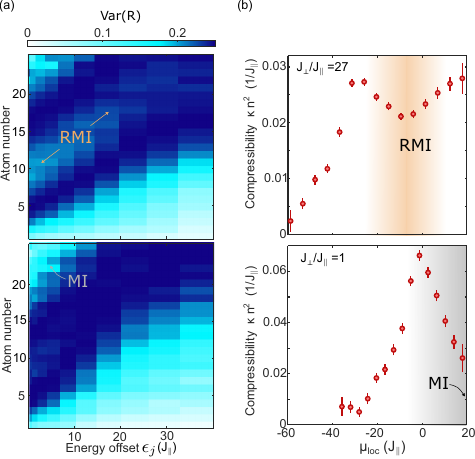}
    \caption{\textbf{Observation of incompressibility in the RMI phase.} (a) Rung parity variance as a function of the atom number and $\epsilon_j$ for $J_\bot/J_\parallel=27$ (upper panel), which shows a region of low variance around 10 atoms that is not present for $J_\bot/J_\parallel=1$ (lower panel). (b) Compressibility as a function of the local chemical potential for $J_\bot/J_\parallel=27$ (upper panel) and $J_\bot/J_\parallel=1$ (lower panel). The shaded regions highlight where the RMI and MI form. 
}
    \label{fig:4}
\end{figure}

Another signature of the RMI is a reduced compressibility, $\kappa = (1/n^2) (\partial n/\partial \mu)$, where $n$ is the site density and $\mu$ is the chemical potential. In the presence of a harmonic trap, the rung-dependent energy offset $\epsilon_{j}$ results in a local chemical potential, $\mu_{loc}=\mu - \epsilon_{j}$. To visualise the effect of $\epsilon_{j}$, we plot the rung parity variance, Var$(R)=R(1-R)$, as a function of $\epsilon_{j}$ and the atom number, for $J_\bot/J_\parallel=1$ and $J_\bot/J_\parallel=27$ [Fig.~\ref{fig:4}(a)]. Both systems have a low variance for atom numbers greater than $20$, but $J_\bot/J_\parallel=27$ has an additional region of low variance, indicating half filling, around 9-12 atoms, where we see the highest $R$ [Fig.~\ref{fig:2}(d)]. 

To calculate the compressibility, we first determine the chemical potential $\mu$ by fitting the density against $\epsilon_j$ using the atom-number probability in the zero-tunnelling limit ($J_\parallel \rightarrow 0$) for a fixed atom number (Supplemental Material, Sec.~\ref{Suppl:Compress}). We then calculate the derivative $\partial n/\partial \mu_{loc}$ to determine the compressibility~\cite{Timon2017} [Fig.~\ref{fig:4}(b, upper panel)], and observe a reduced compressibility of the RMI close to $\mu_{loc}\simeq0$, at half filling ($n=0.5$). This is in contrast to the case $J_\perp/J_\parallel=1$, which shows a maximum compressibility at $\mu_{loc}\simeq0$ and a decrease for higher and lower chemical potentials [Fig.~\ref{fig:4}(b, lower panel)], similar to previous observations in the finite-temperature regime~\cite{sherson2010}.

We have demonstrated the experimental realization of two-leg ladder systems with near half-filling, in which we  prepared and probed the rung-Mott insulator phase. We identified superfluid and insulating regions in the phase diagram using the rung parity inferred from site-resolved occupation measurements. Using amplitude modulation spectroscopy, we probed the excitation spectrum of the RMI and observed a systematic shift of the resonance with respect to the bare many-body gap. This behavior is quantitatively explained by our theoretical modelling, which incorporates the full driven dynamics and local offsets. Our calculations, thus, illustrate how dynamical effects and multiple excited many-body states enter the interpretation of modulation spectroscopy in strongly correlated systems.  Finally, we extracted the compressibility of the RMI  and directly compared it with the Mott insulator, providing further evidence of its incompressible nature at fractional filling.

Our  preparation scheme using programmable light  potentials enables a flexible platform for engineering ladder geometries without the need of a technically challenging superlattice. Beyond the present realization we can extend the system to three-leg ladders and other geometries based on the square lattice, or triangular ladders~\cite{Carfora2026mott}. Another promising direction is the introduction of disorder, enabling the study of localization and glassy behavior in low-dimensional systems, in  comparison with theoretical predictions. This regime has been investigated theoretically for hard-core bosons in two-leg ladders~\cite{Carrasquilla2011,Crepin2011}. Future work could also include the study of transport \cite{li2020localization} and the direct measurement of the phase coherence along the rungs or legs for example via Talbot interferometry \cite{Koehn2025,bruggenjurgen2026phase,su2026quantumcoherencemicroscopehubbard} or quantum gates \cite{impertro2024local}. Such measurements would enable studies of how coherence is restored across insulating-to-superfluid transitions as the system changes from one to two dimensions, and how it is suppressed by disorder.

We thank Lucile Sanchez for feedback on the manuscript. We acknowledge support by the Engineering and Physical Sciences Research Council (EPSRC) through the Quantum Technology Hub for Quantum Computing via Integrated and Interconnected Implementations (QCI3) [EP/Z53318X/1], the Programme Grant ’Quantum Advantage in Quantitative Quantum Simulation’ (QQQS) [EP/Y01510X/1], and the 2020-2021 Doctoral Training Partnership [EP/T517811/1]. TH acknowledges funding from the European Research Council (ERC) under the European Union’s Horizon Europe research and innovation program (Grant Agreement No 101165353 - ERC Starting Grant FOrbQ). 
PK acknowledges funding from EPSRC through grant No. EP/Z533713/1.

\bibliography{LadderRef}

\clearpage

\newpage

\begin{appendices}

\renewcommand\thefigure{S\arabic{figure}}
\renewcommand{\thesubsection}{\Roman{subsection}}

\setcounter{figure}{0}
\label{sec:SI}

\section*{Supplemental Material}

\section{Experimental details}
\subsection{Experimental sequence}

We begin with a Bose-Einstein condensate of approximately $200$ $^{87}$Rb atoms ($F=1,m_F=-1$) confined in an optical dimple trap (wavelength 852\,nm) at the centre of a single anti-node of the vertical lattice (wavelength $\lambda=1064$ nm, depth $V_z=25\Er$, where $\Er=\hbar^2/2m\lambda^2$ is the recoil energy and $m$ is the mass of a \textsuperscript{87}Rb atom). After releasing the atoms from the dimple trap, we prepare a  superfluid in a shallow horizontal 2D lattice ($V=5\Er$ and lattice spacing $\lambda/2=532$ nm). We raise potential barriers in $\unit[200]{ms}$ which creates five isolated two-leg ladder systems. Simultaneously, we adjust $V_{\parallel}$ and $V_{\bot}$ to reach the desired tunnelling amplitudes. To obtain half filling in the centre of the trap, the total atom number is controlled by varying the depth of the dimple trap during evaporative cooling. Site-resolved images are obtained by collecting fluorescence photons with a high-numerical-aperture microscope objective. Owing to light-assisted collisions during imaging, the  quantum-gas microscope detects  the parity of the site occupation, such that doubly-occupied sites appear empty~\cite{sherson2010}. The parity of the lattice occupation is reconstructed using the  Lucy-Richardson deconvolution algorithm~\cite{larooij2022}.

\subsection{Experimental values and temperatures}

Throughout the paper, we express quantities in terms of $J_\perp$, $J_\parallel$ and $U$, which all change as a function of lattice depth. Here we list the relevant Hubbard parameters at half-filling for the datasets shown in Fig.\,\ref{fig:3}. We also show the temperature in units of $J_\parallel$, and the energy gap, $\Delta E/J_\parallel$. The temperature  corresponds to the
centre of the uncertainty band in Fig.~\ref{fig:2}(b). The energy gap results from  the DMRG calculation from Fig.~\ref{fig:2}(a).

\begin{table}[b]
    \centering
    \color{black}
    \renewcommand{\arraystretch}{1.5}
    \setlength{\tabcolsep}{4pt} 
        \begin{tabular}{ccccccc}
        \hline
\!\!$V_\parallel(\!E_r)$ &
\!\!$V_\perp(\!E_r)$ &
$U(\mathrm{Hz})$ &
$J_\parallel(\mathrm{Hz})$ &
$J_\perp(\mathrm{Hz})$ &
$T/J_\parallel$ &
$\Delta E/J_\parallel$ \\
\hline
        20 & 5 & 548 & 5 & 134 & 12(2) & 27.0\\
        20 & 7 & 615 & 5 & 80 & 12(2) & 19.9\\
        20 & 9 & 673 & 5 & 49 & 12(2) & 13.0\\
        20 & 11 & 717 & 5 & 31 & 12(2) & 7.9\\
        20 & 13 & 754 & 5 & 20 & 12(2) & 4.5\\
    \hline
    \end{tabular}
\caption{Experimental values and temperatures for the datasets in Fig.\,\ref{fig:3}.}
    \label{tab:example}    
\end{table}

For the last two datasets in the table (two left most datapoints in Fig.\,\ref{fig:3}) the energy gap is smaller than the temperature. Nevertheless, we still observe a reduction in  rung parity in the modulation measurements. This can be understood from the fact that the modulation spectroscopy probes the energy  gaps in the excitation  spectrum, through transitions induced by the modulation, rather than directly measuring the thermal populations. Consequently, a temperature exceeding the many-body gap does not necessarily wash out the spectroscopic response associated with this gap.

\subsection{Programmable light potentials}
\label{S:light}
To create the repulsive potential barriers,  we employ  a  digital mirror device (DMD, model ViALUX V-9001) illuminated with   blue-detuned light at a wavelength of \unit[666]{nm} from a single-mode amplified diode laser \cite{DiCarli2024Commensurate}. The light potential is projected onto the atoms by the high-resolution objective, with a  demagnification of the imaging system  such that each lattice site corresponds to approximately $18\times 18$ DMD pixels. The barrier height is made spatially uniform using an intensity-feedback algorithm~\cite{Koehn2025}, and is set to approximately $3U$ in the RMI regime. {This value was chosen such that the probability of finding  an atom on a barrier during the sample preparation is negligible. However, uncertainties in the barrier height  $(<5\%)$ can create a site-to-site potential disorder. We estimate that at maximum 5\% of the repulsive light potential from the barriers leaks onto the neighboring ladder sites. This corresponds to a disorder strength $\Delta\simeq2.5J_\parallel=h\times 12.5$ Hz for $V_\parallel=20\Er$, which is less than 10\% of the energy gap $\Delta E=27 J_\parallel$    at $V_\parallel=20\Er,V_\perp=5 \Er$. We have defined  the disorder strength, $\Delta$, as the corresponding maximum variation in the local chemical potential \cite{Koehn2025}}. 

In our experiment, the position on the light potential drifts relative to the lattice sites of the  horizontal lattice  due to  thermal effects, amounting to typically 1-4 lattice sites per hour. To compensate for this drift, we determine the phase of the horizontal lattices from  the  the positions of individual  atoms in the fluorescence image, and  adjust the light potential accordingly by translating the potential pattern on the DMD by a few pixels  \cite{DiCarli2024Commensurate,Koehn2025}. The lattice-phase measurement has an uncertainty of approximately $5\%$ of the lattice spacing.

\subsection{Filling fraction and imbalance} 
\begin{figure}[!t]
    \centering
    \includegraphics[width=0.8\linewidth]{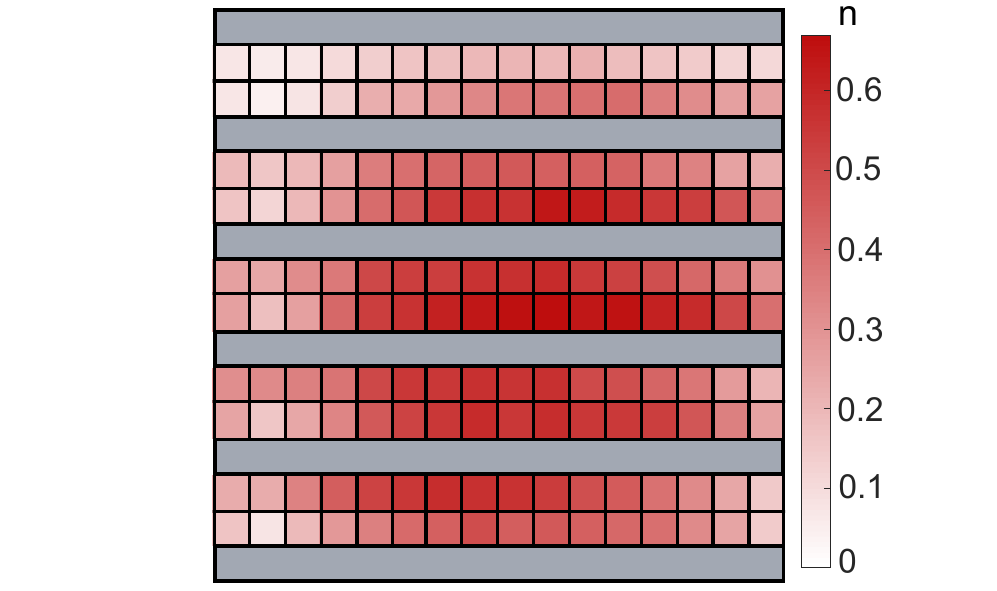}
    \caption{\textbf{Site occupations.} 
    Measured mean  occupation of each lattice site in the five two-leg ladders, averaged over 
    $\sim 1600$ experimental realisations using the same datasets as Fig.~\ref{fig:4}(a, upper panel). The grey squares show the programmable light potentials, which form repulsive barriers between the ladders. Because of the harmonic confinement, the two legs of each outer ladder have a potential bias $\epsilon_B\simeq0.23J_\perp$. This offset is $\sim20\%$ of the maximum bias resulting from the positioning errors of the projected light potentials.
      }
    \label{fig:S1}
\end{figure}

We determine the average site occupations $n_{i,j}$ for each lattice site across the different ladder systems. To illustrate the spatial occupation profile, we use the same datasets as in the analysis of  Fig.~\ref{fig:4}(a, upper panel), and plot the site-resolved mean occupations in Fig.~\ref{fig:S1}. From this, we compute the imbalance between the two legs,
    $I=\frac{n_{1,j}-n_{2,j}}{n_{1,j}+n_{2,j}}$.
This imbalance is $<0.1$ for the center ladders, and reaches $0.35$ for the outer ladders. This behavior is  primarily caused by the harmonic confinement. The vertical lattice provides the dominant  contribution to the harmonic confinement for which we measured a trapping frequency $\Omega_z=2\pi \times 42(2)$Hz for $V_z=25\Er$ and beam waist $\omega_0=130(3)\,\textmu$m. The curvature of the trap is $\frac{\partial^2 \epsilon}{\partial x^2}\simeq1.2 J_\parallel$ for $V_\parallel=20\Er$. In addition, there are potential offsets caused by errors in the placement of the light potentials that vary between experimental runs. However, these are not visible in the averaged site occupations.  For the data shown  in Fig.\,\ref{fig:2}, after post-selection, each of the five ladders contribute a similar amount of data sets. 

To quantify the degree of delocalization over the rung, we  calculated with exact diagonalization the two-site Bose-Hubbard model for a single atom,  to infer the probabilities $P_L$ and $P_R$, to find the atom in the left and right sites, respectively, for different bias potentials $\epsilon_B$, and $V_\perp=5\Er$ (see Table II). 
The ground-state wavefunction has an increased probability at the site without offset.  

\begin{table}[!h]
    \centering
    \renewcommand{\arraystretch}{1.5}
        \setlength{\tabcolsep}{6pt} 
    \label{tab:example}
    \begin{tabular}{ccccc}
        \hline
        $\epsilon_B (\rm{Hz})$ & $\epsilon_B (J_\perp)$ & $P_L$ & $P_R$ \\
        \hline
        0 & 0 & 0.50 & 0.50 \\
        50 & 0.37 & 0.59 & 0.41 \\
        100 & 0.75 & 0.67 & 0.33 \\
        150 & 1.12 & 0.74 & 0.26 \\
        \hline
    \end{tabular}
    \caption{Population imbalance for different bias potential offsets resulting from exact diagonalization.}

\end{table}

\FloatBarrier

\subsection{Analysis of compressibility measurements}
\label{Suppl:Compress}
\begin{figure}[!t]
    \centering
    \includegraphics[width=1\linewidth]{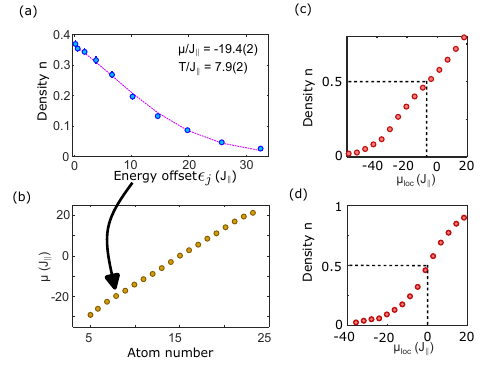}
    \caption{\textbf{Compressibility measurement.} 
      (a) Fit of  the density and potential offset for the 8-atom dataset, with  temperature and chemical potential as free parameters. (b) Chemical potential, extracted  from the fits for each atom number   for atom numbers 5 to 23. (c) Density as a function of  the local chemical potential in the  RMI regime. (d) Density as a function of the local chemical potential in the case $J_\perp/J_\parallel=1$. The dashed black line highlights half filling ($n=0.5$). Error bars are smaller than most of the data points.}
    \label{Fig.S2}
\end{figure}

To calculate the local particle-number compressibility of the ladder systems [Fig.~\ref{fig:4}(b)], we first determine  the energy offset induced by  the global harmonic trap for each ladder. The trap center is identified from the center-of-mass of the atomic cloud,  and the local energy offset from the harmonic trap, $\epsilon_{i,j}$, is evaluated as a function of the distance $r$ in lattice sites from this center. For each ladder, the potential of the two legs are averaged to obtain the rung-dependent energy offset $\epsilon_j$. We group the ladders by observed atom number , using the range $5-23$, and calculate the density as a function of $\epsilon_j$ for each atom number [Fig.~\ref{Fig.S2}(a)].

The next step is to fit this density to determine the global chemical potential, $\mu$, and temperature, $T$, for each atom number. In the zero-tunnelling limit  ($J_\parallel \rightarrow 0$), each rung is treated as an isolated two-site system with eigenstates, $k$, eigenenergies, $E_k$, and occupations, $N_k$. The probability of observing a rung with occupation $n$ is obtained from the grand-canonical ensemble by summing over all eigenstates with $N_k=n$, $P_{n}(j)= e^{\beta[\mu_{\rm loc}(j)N_k-E_k]}/Z(j)$, where $\beta=1/(k_BT)$ and $Z(j)=\sum_k e^{\beta[\mu_{\rm loc}(j)N_k-E_k]}$ is the local partition function, and $k_B$ is the Boltzmann constant. The energy spectrum $E_k$ is obtained from exact diagonalization of a single rung and includes all states with $N_k=0, 1$ and $2$ atoms. Specifically, we include the state corresponding to $N=0,E_0=0$, the symmetric and anti-symmetric single-particle states ($N=1,E_1^+= - J_\bot$ and $E_1^-= + J_\bot$) and the relevant two-particle eigenstates for $N=2$, including interaction $U$. The global chemical potential $\mu$ and temperature $T$ are the only free parameters in the fit. The procedure is  repeated for each atom number, yielding the plot shown in Fig.~\ref{Fig.S2}(b).
Using the fitted values of $\mu$, we calculate the local chemical potential $\mu_{loc}=\mu-\epsilon_{j}$, which  allows us to show the density profiles as a function of $\mu_{loc}$ in a single graph [Figs.~\ref{Fig.S2}(c) and \ref{Fig.S2}(d)]. These datasets are then subsequently binned and differentiated with respect to $\mu_{\rm loc}$ to determine the compressibility shown in Figs.~\ref{fig:4}(b) and \ref{fig:4}(c).

\section{Numerical simulations}\label{suppl:Simulations}

In this section we describe the methods used to perform the numerical simulations of the experimental system. We note that many of the methods and results presented here are based on those introduced in Ref.~\cite{Carfora2026mott}.

\subsection{Ground-state calculations }
To determine the low-energy spectrum of the Bose-Hubbard Hamiltonian described by Eq.~\eqref{eq:BHM}, we employed the standard DMRG algorithm~\cite{schollwock2005density, schollwock2011density, orus2019tensor}, as implemented in the \textsc{ITensor} library~\cite{Itensor1, Itensor2}. 
The algorithm returns the ground-state and first excited-state energies, $E_0$ and $E_1$, together with the corresponding matrix product states representing the respective eigenstates and energy gap $\Delta E=E_1-E_0$. Using these results, we obtained  the phase diagram of a $L=N=10$ ladder lattice, as shown in  Fig.~\ref{fig:2}(a). 
The Hamiltonian parameters, $J_\parallel$, $J_\perp$ and $U$, were calculated numerically from the experimental lattice depths $V_\parallel$ and $V_\perp$~\cite{Lewenstein2012}. The simulations use open boundary conditions and neglect the harmonic confinement, because in the RMI phase of a system of size $L=10$, the energy shift due to the harmonic confinement at the outermost sites is smaller than the energy gap.

The required matrix product operators were constructed by labelling lattice sites sequentially along the rungs. The DMRG algorithm contracts pairs of neighbouring tensors describing the singular sites. After optimizing said contractions, the site tensors are re-obtained via single value decomposition. The following process is done over all pairs of neighbouring site tensors, sweeping from one end of the chain to the other \cite{white1992dmrg, schollwock2011density}. To obtain the ground (first excited) state, 150 (250) sweeps were performed. Singular values smaller than $10^{-8}$ were discarded, and the maximum local occupation number was restricted to 6 particles per site. 

The numerical results of Figs.~\ref{fig:2}(b) and \ref{fig:2}(c) are similar to those reported in Ref.~\cite{Carfora2026mott} while additionally including the effect of finite temperature. We followed the purification approach described  in Ref.~\cite{schollwock2005density}, in which an ancilla site is introduced for each physical site in the system. This means that our simulations require double the number of sites while the Hamiltonian acts only on the physical sites, and the maximum system size accessible in these simulations is approximately reduced by a factor of two. We then construct a maximally entangled state between each system site and its ancilla. Tracing out the ancilla degrees of freedom of said state represents the density matrix at infinite temperature. To ensure that the total number of particles is conserved, we use the $N$-particle state~\cite{barthel2016canonical}
\begin{equation}\label{eq:inft_canonical}
    |\psi_{\beta=0}^N\rangle= \frac{(\mathcal{A})^N}{\sqrt{N!}}\ket{\text{vac}} , \hspace{0.4cm} \mathcal{A}=\sum_ia^\dagger_{i,P}a^\dagger_{i,A},
\end{equation}
where $\ket{\text{vac}}$ denotes the vacuum state for both the physical and ancilla sites. The operator $\mathcal{A}$ uses the creation operators $a^\dagger_{i,P}$ and $a^\dagger_{i,A}$  acting on the $i^\text{th}$ physical and ancilla sites, respectively. 
We obtain finite-temperature thermal states by performing imaginary-time evolution using the TEBD algorithm~\cite{PAECKEL2019167998}. We limit the calculations to a maximum on-site occupation of $4$ bosons, and obtained solutions for $\beta=30/E_\text{r}$ and $\beta=40/E_\text{r}$ for a half-filled ladder with $L=8$ rungs.

\subsection{Lattice modulation}

\begin{figure}[!b]
    \centering
    \includegraphics[width=0.9\linewidth]{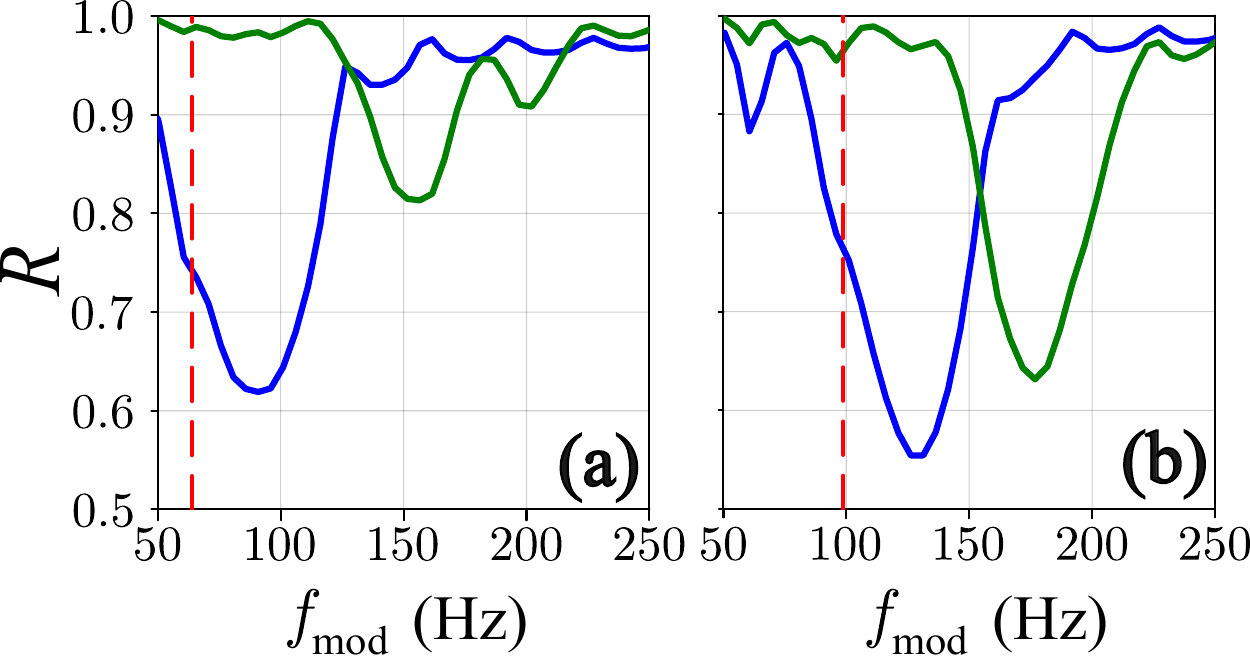}
    \caption{\textbf{Lattice modulation simulations.} Average rung parity after frequency modulation for $t=30$~ms for a ladder with $L=N=16$ rungs. Results are shown for $J_\parallel/h\approx5$ Hz ($V_\parallel=20 E_\text{r}$), with (a) $J_\perp/J_\parallel\approx10$ ($V_\perp=9E_\text{r}$) and (b) $J_\perp/J_\parallel\approx16$ ($V_\perp=7E_\text{r}$). Results for both no offset between the two chains ($\epsilon_B=0$,  blue lines) and the maximum offset shown in Fig.~\ref{fig:3} of the main text ($\epsilon_B=1.1J_\perp$, green lines) are shown.
    The red dashed lines are the energy gaps predicted by the DMRG method, which are (a) $\Delta E\approx 0.75J_\perp$ and (b) $\Delta E\approx0.49J_\perp$.}

    \label{fig:freq_mod}
\end{figure}

In Fig.~\ref{fig:3}, we compare the experimentally observed lattice modulation resonance frequency with predictions from  both the energy gap of the system and direct numerical simulations of the experimental lattice modulation. The lattice modulation is implemented  such that the lattice depth along the chains is time-dependent,
\begin{equation}\label{eq3}
V_\parallel(t)=V_{\parallel}[1+A\sin(2\pi f_\text{mod}t)],
\end{equation}
where $V_{\parallel}$ is the initial lattice depth before the modulation is applied and $A$ is the dimensionless modulation amplitude. For the experimental results presented in the main text, we used $A=1/4$. The time-dependent Bose-Hubbard parameters $U(t)$, $J_\parallel(t)$ and $J_\perp$ are calculated from the lattice depth $V_\parallel(t)$ and the fixed $V_\perp$, with $U(0)=U$ and $J_\parallel(0)=J_\parallel$. The ground state is obtained via DMRG and used as the initial state, and evolved for a duration of 30~ms using TEBD under the time-dependent Hamiltonian, $H(t)=U(t)H_U-J_\parallel(t)H_\parallel-J_\perp H_\perp$.
After the evolution, we calculate the rung parity $R$ and plot it as a function of the modulation frequency $f_\text{mod}$.

The simulations were performed for several lattice depths, corresponding to different tunnelling rates (Fig.~\ref{fig:freq_mod}). The numerical results show a clear reduction in rung parity, however both the experimental measurements and the simulations show a  resonance  at a higher frequency  than  the energy gap obtained from  the DMRG simulations. This occurs because the modulation couples not only to the lowest excited state but also to higher-energy states, as discussed  in detail in the following section \ref{Suppl:modulation}. Although this coupling accounts for part of the observed shift, the displacement of the resonance in the numerical results remains smaller than that measured experimentally.  We attribute this remaining discrepancy  to experimental imperfections, in particular the  presence of a small potential bias between the two legs of the ladder. The origin and effect of this bias is described in detail in Sec.~\ref{suppl:bias}.

\subsubsection{Power broadening and transition strengths}

\label{Suppl:modulation}

\begin{figure}[!b]
    \centering
    \includegraphics[width=0.8\linewidth]{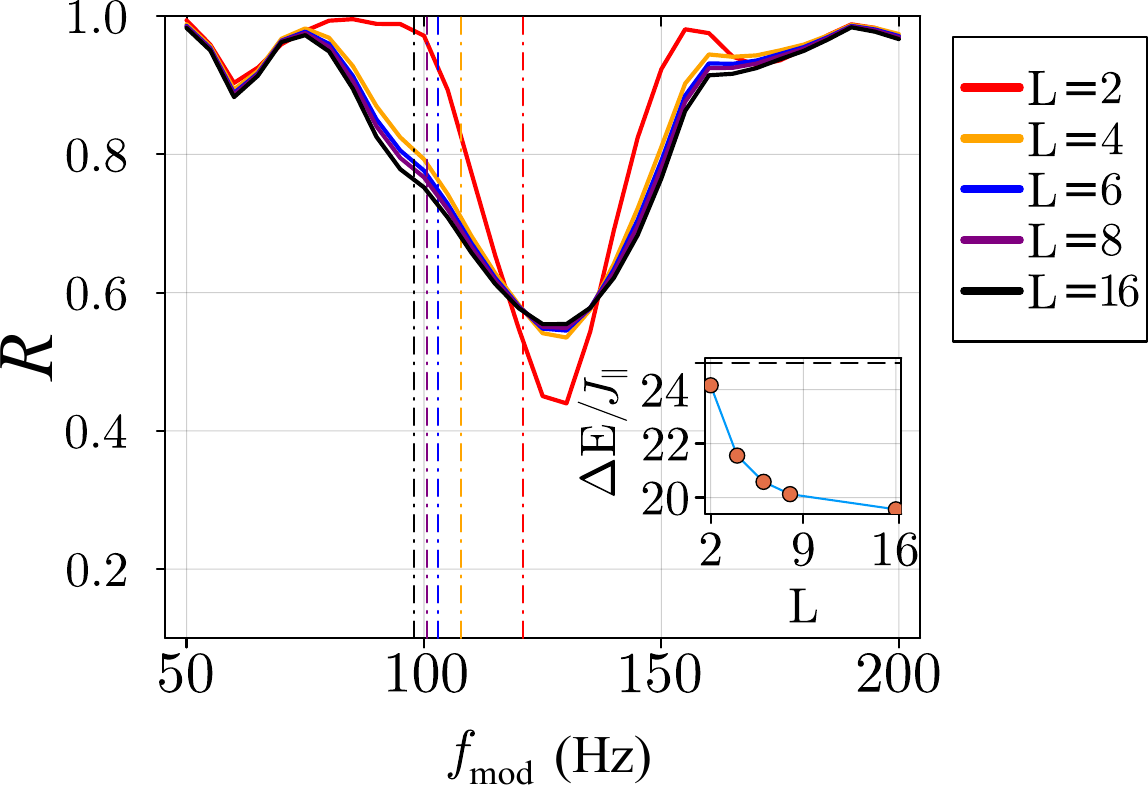}
    \caption{\textbf{System size scaling.} Frequency modulation for $J_\parallel/h\approx5$ Hz ($V_\parallel=20E_\text{r}$), $J_\perp/J_\parallel\approx16$ ($V_\perp=7E_\text{r}$) for different number of rungs $L$. The dashed lines show the energy gaps of the respective system sizes. The inset shows  how the size of the gap changes with respect to $L$.}
\label{fig:scaling_modulation}
\end{figure}

\begin{figure}[!t]
    \centering
    \includegraphics[width=0.99\linewidth]{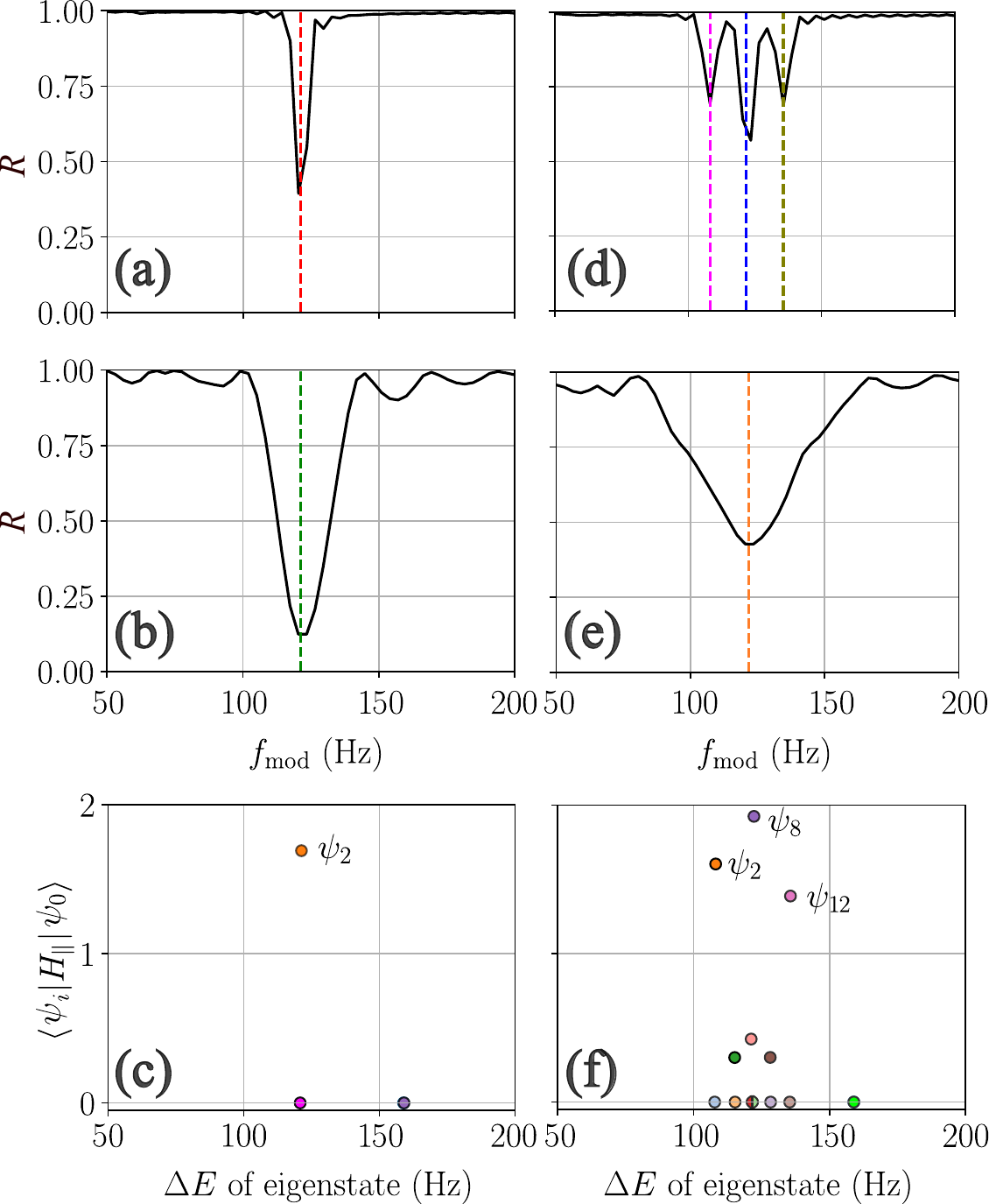}
    \caption{\textbf{Frequency modulation for small systems.} Numerical results for $V_\parallel=20E_\text{r}$, $V_\perp=7E_\text{r}$ of the frequency modulation for $L=2$ (a)--(c) and $L=4$ (d)--(f) for weak driving [$A=1/16$, in panels (a) and (d)] and for strong driving [$A=1/4$, in panels (b) and (e)]. All results are taken after evolving for $t\hbar A/J_\parallel=0.32$. The matrix elements of the $H_\parallel$ component between the ground state $\psi_0$ and the excited states $\psi_i$ are shown as a function of the corresponding excitation energy $\Delta E_i=E_i-E_0$ in panels (c) for $L=2$ and (f) for $L=4$. The dominant matrix elements are labelled by the corresponding excited $\psi_i$ states that are coupled to  $\psi_0$. The $\langle\psi_i|H_\parallel|\psi_0\rangle$ elements in panel (c) are colour-coded with the corresponding populations $|\langle\psi_i(t)|\psi_i(0)\rangle|^2$ in Fig.~\ref{fig:time_evol_L2}, while the elements in panel (f) with the populations in Fig.~\ref{fig:time_evol_L4}.}
    \label{fig:freq_mod_size_amp}
\end{figure}

There is a systematic shift between  the resonance positions obtained from the full time-dependent simulations and the energy gaps, $\Delta E$, calculated using the DMRG method. This discrepancy is primarily caused by power broadening due to the finite-strength  drive. For a two-rung system  ($L=2$), the energy gap lies  close to the centre of the resonance. However, as the system size increases, the simulated energy gap shifts values,  while the resonance frequency does not shift significantly (Fig.~\ref{fig:scaling_modulation}). To investigate this effect, we consider simplified  models of  only $L=2$ or $L=4$ rungs, for which  exact diagonalization can be performed  away from the hard-core boson  limit. We consider weak ($A=1/16$) and strong ($A=1/4$) modulations, and  show the resulting modulation curves for $L=2$ at half filling, $N=2$, in Figs.~\ref{fig:freq_mod_size_amp}(a) and \ref{fig:freq_mod_size_amp}(b). The width of the resonance becomes  broader as the driving strength is increased towards the value used in the experiment. The position of the resonance  in this case is very close to the energy gap, consistent with the results of the scaling analysis. 

\begin{figure}[!t]
    \centering
    \includegraphics[width=0.85\linewidth]{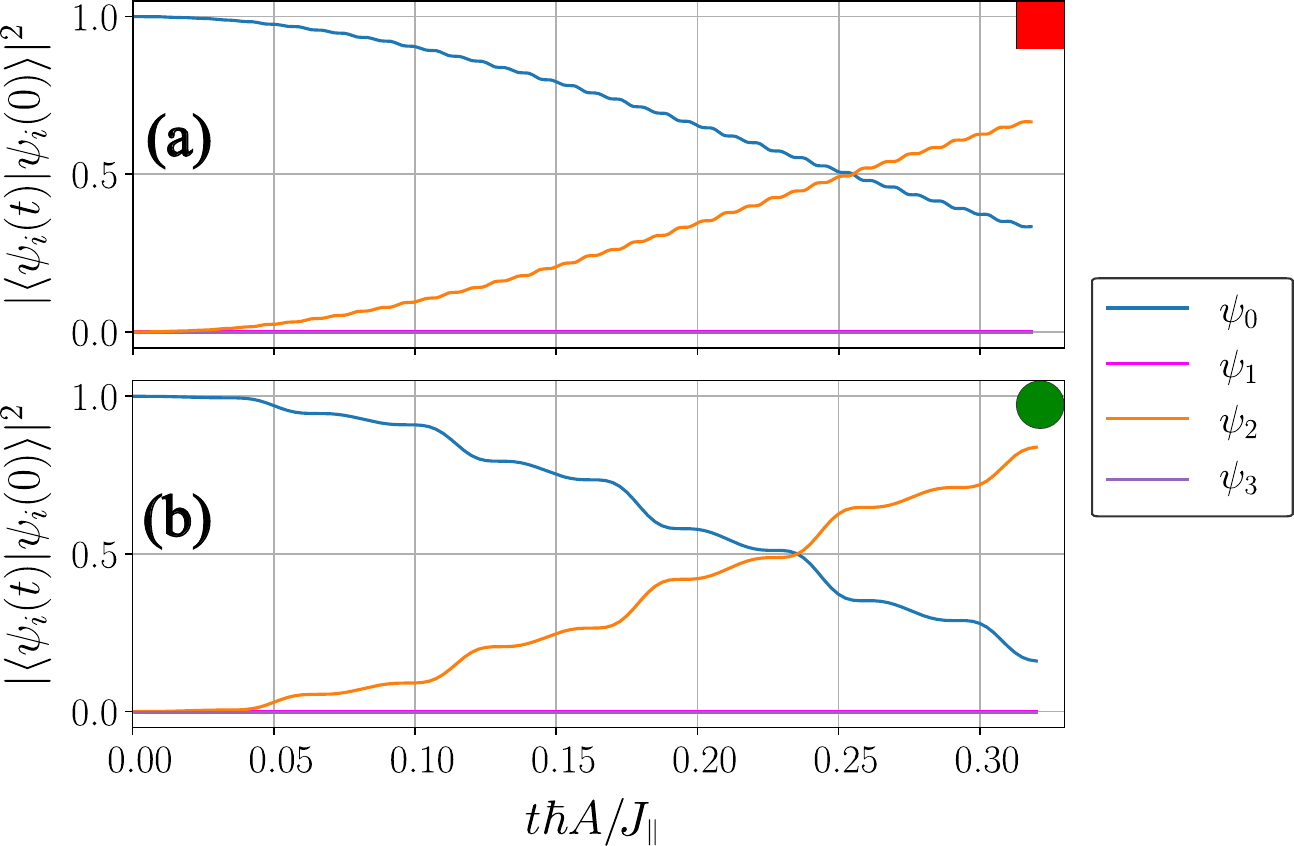}
    \caption{\textbf{Population dynamics for $\mathbf{L=N=2}$}. Time evolution of the eigenstate populations at resonance for $V_\parallel=20E_\text{r}$, $V_\perp=7E_\text{r}$, for (a)  $A=1/16$ and (b) $A=1/4$. Time is expressed in units of  $J_\parallel$ at the start of the modulation. The value of $f_\text{mod}$ is indicated by dashed lines in Figs.~\ref{fig:freq_mod_size_amp}(a) and \ref{fig:freq_mod_size_amp}(b) using the same colour as the markers  in the top-right corner of the panels (square for $A=1/16$, circle for $A=1/4$). In both cases, the ground $\psi_0$ state is coupled only to $\psi_2$. 
    \label{fig:time_evol_L2}}
\end{figure}
 
 The origin of this shift can be understood by considering a larger system with $L=4$ rungs, as shown in Figs.~\ref{fig:freq_mod_size_amp}(d)--(f). For weak driving, the modulation spectrum shows multiple distinct resonances, each corresponding to a transition from the ground state to a different excited eigenstate. The accessible excited states are illustrated in Figs.~\ref{fig:freq_mod_size_amp}(c) and (f), where we also show the relevant matrix elements for the coupling between the ground state and each excited state. These individual transitions are resolved for weak driving, $A=1/16$, but broaden and merge into a single feature as the modulation  strength is increased. As a result, the centre of the broadened resonance shifts towards a frequency higher than that of the lowest-energy excitation. Notably, the resulting resonance frequency is close to the energy gap of the corresponding $L=2$ system.

\begin{figure}[!t]
    \centering
    \includegraphics[width=0.85\linewidth]{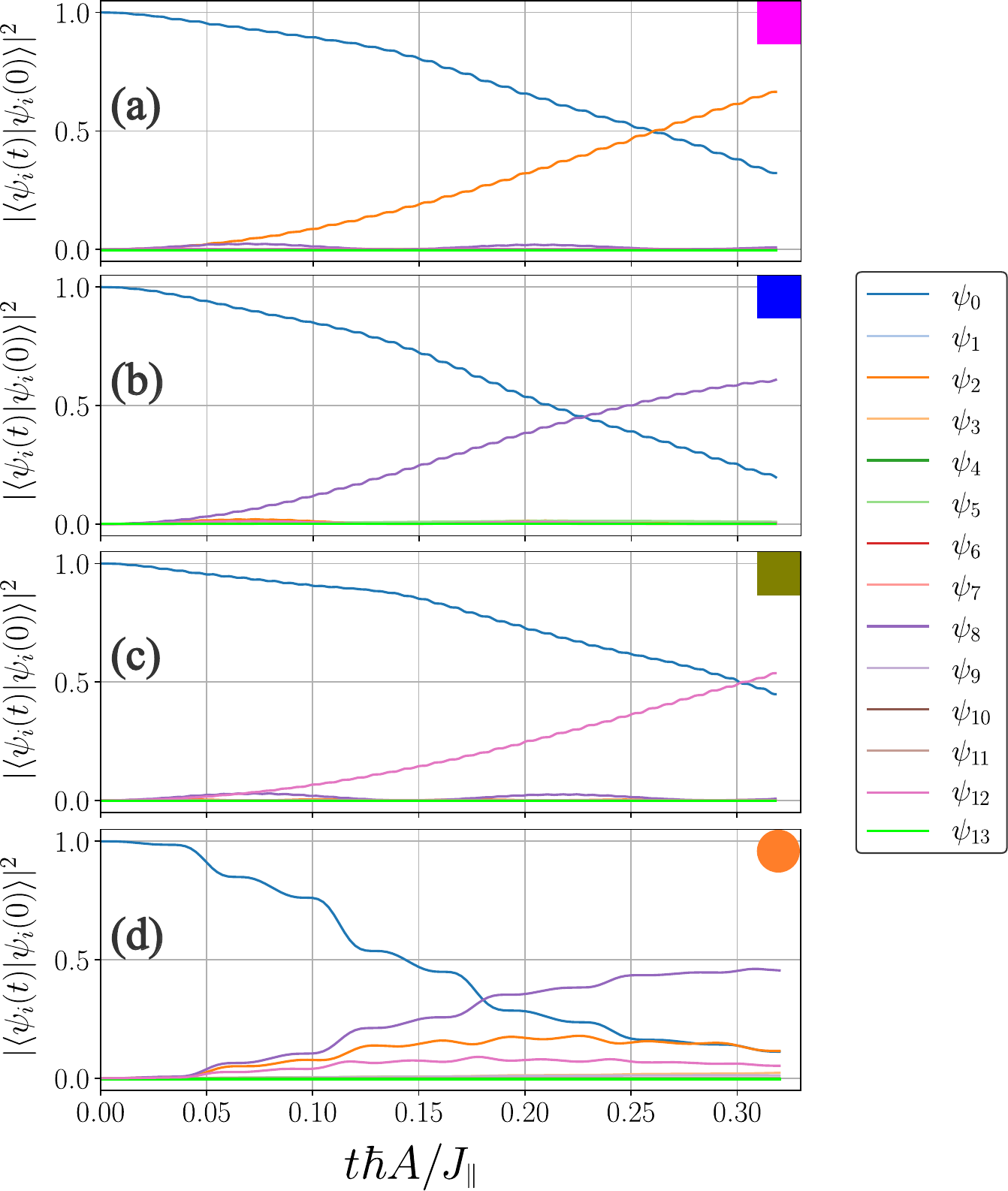}
    \caption{\textbf{Population dynamics for $\mathbf{L=N=4}$}. 
    (a)--(c) Time evolution of the eigenstate populations for $A=1/16$ for shaking at different resonance frequencies, with $V_\parallel=20E_\text{r}$, $V_\perp=7E_\text{r}$. The corresponding resonance frequencies are identified  by the dashed lines in the spectra of Figs.\ref{fig:freq_mod_size_amp}(d) and \ref{fig:freq_mod_size_amp}(e)  with the same colour as the markers shown in the top right corner of each panel (square for $A=1/16$, circle for $A=1/4$). Time is expressed in units of  $J_\parallel$ defined at the start of  the modulation. (d) Rung parity dynamics for $A=1/4$, with the modulation frequency set to the centre of the broadened resonance shown in Fig.~\ref{fig:freq_mod_size_amp}(e).
    }
    \label{fig:time_evol_L4}
\end{figure}

To identify  which states are involved in these transitions, we show the dynamics of the population of the ground state $\psi_0$ and excited states $\psi_i$. For $L=2$, only the second excited state, $\psi_2$, becomes populated (Fig.~\ref{fig:time_evol_L2}). The first excited state, $\psi_1$, is a dark state that is not coupled to the ground state by the driving Hamiltonian.
However, the energies of the dark and bright states are so close to each other that there is no shift in the location of the resonance. This is also evident from the matrix elements, as $\bra{\psi_1} H_\parallel\ket{\psi_0}$ vanishes, whereas  $\bra{\psi_2} H_\parallel\ket{\psi_0}$ is large [Fig.~\ref{fig:freq_mod_size_amp}(c)]. Apart from differences in the detailed population-transfer dynamics, the behaviour is qualitatively similar for strong and weak driving (Fig.~\ref{fig:time_evol_L2}).

We now consider $L=4$ to investigate how this mechanism extends to longer ladders (Fig.~\ref{fig:time_evol_L4}). For weak driving, we show the population dynamics at each resonant frequency shown in Fig.~\ref{fig:freq_mod_size_amp}(d). These resonances correspond to transitions from $\psi_0$ to the states  $\psi_2$, $\psi_8$ and $\psi_{12}$ respectively, that have  the largest transition  matrix elements [Fig.~\ref{fig:freq_mod_size_amp}(f)]. For strong driving,  these transitions are simultaneously  excited, causing the individual resonances to overlap, and a shift of  the resonance frequency  towards higher values.
 
\FloatBarrier

\subsubsection{Light-potential bias }

\label{suppl:bias}
We now consider the effect of a small potential bias, $\epsilon_B$, between the two sites of each rung. Such a bias can arise due to imperfections in the alignment of the projected potential walls relative to the lattice sites. These imperfections originate both from the uncertainty in determining the lattice phase (see Section \ref{S:light}), and from mechanical instabilities of the optical setup,  which causes the position of the projected light potential to  vary slightly between different realisations of the experiment. These shifts correspond to only a small fraction of a lattice spacing at the position of the atoms. We incorporate this energy offset into the numerical simulations and find that it shifts the resonance to higher frequencies. To quantify this effect, we calculated the frequency resonance of a $L=N=16$ ladder with $\epsilon_B=0.56J_\perp$ and $\epsilon_B=1.1J_\perp$. The results define the boundaries of the red shaded region shown in Fig.~\ref{fig:3}. 
Further examples illustrating the effect of the bias on the frequency-modulation spectra are shown in Fig.~\ref{fig:freq_mod}. The energy offset induces an effective perpendicular hopping rate $\tilde{J}_\perp\approx\sqrt{J_\perp^2+\epsilon_B^2}$, and leads to a less defined resonance for weaker values of $J_\perp/J_\parallel$ (or equivalently, for higher values of $V_\perp$). This behavior is expected in the limit of $\epsilon_B\gg U,J_\parallel,J_\perp$. In this case, the system effectively reduces to a fully-filled 1D chain whose dynamics are insensitive to the modulation. However, for all values of  $J_\perp/J_\parallel$ considered in the experiment, the atoms remain delocalised over the two sites of each rung despite the potential bias.

\end{appendices}
\end{document}